\documentclass[12pt]{article}
\usepackage{amsmath,amssymb}
\newcommand{\eqdef}{\stackrel{\text{def}}{=}}
\newcommand{\n}{\nonumber \\}
\renewcommand{\theequation}{\arabic{section}.\arabic{equation}}

\allowdisplaybreaks[3]

\begin{document}

\baselineskip=20pt

\newfont{\elevenmib}{cmmib10 scaled\magstep1}
\newcommand{\preprint}{
     \begin{flushleft}
       \elevenmib Yukawa\, Institute\, Kyoto\\
     \end{flushleft}\vspace{-1.3cm}
     \begin{flushright}\normalsize  \sf
       DPSU-07-1\\
       YITP-07-26\\
       {\tt arXiv:0706.0768}\\
       June 2007
     \end{flushright}}
\newcommand{\Title}[1]{{\baselineskip=26pt
     \begin{center} \Large \bf #1 \\ \ \\ \end{center}}}
\newcommand{\Author}{\begin{center}
     \large \bf Satoru~Odake${}^a$ and Ryu~Sasaki${}^b$ \end{center}}
\newcommand{\Address}{\begin{center}
       $^a$ Department of Physics, Shinshu University,\\
       Matsumoto 390-8621, Japan\\
       ${}^b$ Yukawa Institute for Theoretical Physics,\\
       Kyoto University, Kyoto 606-8502, Japan
     \end{center}}
\newcommand{\Accepted}[1]{\begin{center}
     {\large \sf #1}\\ \vspace{1mm}{\small \sf
Accepted for Publication}
     \end{center}}

\preprint
\thispagestyle{empty}
\bigskip\bigskip\bigskip

\Title{Exact Heisenberg operator solutions for
multi-particle quantum mechanics}
\Author

\Address

\begin{abstract}
Exact Heisenberg operator solutions for independent `sinusoidal
coordinates' as many as the degree of freedom are derived for typical
exactly solvable multi-particle quantum mechanical systems, the Calogero
systems based on any root system.
These Heisenberg operator solutions also present the explicit forms of
the {\em annihilation-creation operators\/} for various quanta in the
interacting multi-particle systems.
At the same time they can be interpreted as {\em multi-variable
generalisation\/} of the {\em three term recursion relations\/}
for multi-variable orthogonal polynomials constituting the eigenfunctions.
\end{abstract}

\section{Introduction}
\label{intro}
\setcounter{equation}{0}

Modern quantum physics is virtually unthinkable without
annihilation-creation operators, which are defined as the
positive/negative energy parts of the {\em Heisenberg field operator
solutions\/} of a {\em free\/} field theory, an infinite collection of
independent harmonic oscillators. These annihilation-creation operators
map an eigenstate of a free Hamiltonian into another, but not connecting
those of a full theory.
Our knowledge of the Heisenberg operator solutions of a full interacting
theory, on the other hand, is quite limited in spite of the central role
played by the Heisenberg operator solutions in field theory in general.
In the so-called exactly solvable quantum field theories, factorised
$S$-matrices and some of the correlation (Green's) functions are the 
highest achieved points up to now.

Following our embryonic work \cite{os7} on the construction of exact 
Heisenberg operator solutions for various degree one quantum mechanics,
we present in this paper a modest first step in the quest of deriving
exact Heisenberg operator solutions for a family of interacting
multi-particle dynamics. These are the Calogero systems \cite{cal},
which are integrable multi-particle dynamics based on root systems. 
For the theories based on the classical root systems, the $A$, $B$, $C$
and $D$ series, the number of particles can be as large as wanted,
but not infinite as in field theories.
A complete set of exact Heisenberg solutions for `sinusoidal coordinates'
\cite{os7,nieto}, as many as the degree of freedom, is derived
elementarily in terms of the universal Lax pair \cite{bcs1,bms,kps},
which is a well-established solution mechanism for classical and quantum
Calogero-Sutherland-Moser systems \cite{cal,sut,mos} based on any root
system \cite{OP1}.
Explicit forms of various annihilation-creation operators are obtained 
as the positive/negative energy parts of the Heisenberg operator solutions.
They map an eigenvector of the full Hamiltonian into another.
These sinusoidal coordinates and the corresponding annihilation-creation
operators provide multi-variable generalisation of the three term recursion 
relations \cite{os7} of orthogonal polynomials constituting the
eigenfunctions of the Calogero Hamiltonian.

This paper is organised as follows.
In section two, the rudimentary facts and notation of Calogero systems
based on any root system are recapitulated together with the universal
Lax pair matrices.
In section three the complete set of exact Heisenberg operator solutions
is derived quite elementarily based on generating functions constructed
from the universal Lax matrices.
Remarks on the special features of the $D$-type theories are given at
the end of the section.
The final section is for a summary and comments for further research.
The Appendix gives the list of the preferred sets of weight vectors for
explicit representations of Lax matrices based on the exceptional and the
non-crystallographic root systems.

\section{Calogero Systems}
\label{calogero}
\setcounter{equation}{0}

In this paper we will derive {\em exact Heisenberg operator solutions
for the Calogero systems\/}. They are  one-dimensional multi-particle 
dynamics with inverse (distance)${}^2$ potential inside a harmonic
confining potential.
They have a remarkable property that they are exactly solvable at the
classical \cite{st2} and quantum \cite{cal,kps} levels.
The exact quantum solvability has been shown in the {\em Schr\"odinger
picture\/}, as the entire energy spectrum is known and the corresponding
eigenfunctions can be constructed explicitly by a finite number of
algebraic processes (the lower triangularity of the Hamiltonian in
certain basis) \cite{kps}.
The exact quantum solvability in the {\em Heisenberg picture\/} of the
Calogero systems will be demonstrated in the next section by constructing
the explicit Heisenberg operator solutions for the independent
`sinusoidal coordinates' as many as the degree of freedom.
Let us briefly recapitulate the essence of the quantum Calogero systems
\cite{cal,kps} together with appropriate notation necessary in this paper.

\subsection{Quantum Hamiltonian}
\label{quanthamsec}

A Calogero system is a Hamiltonian dynamics associated with a root
system \cite{bcs1,OP1} $\Delta$ of rank $r$, which is a set of
vectors in $\mathbb{R}^r$ with its standard inner product. Its
dynamical variables are the coordinates $\{q_{j}\}$ and their
canonically conjugate momenta $\{p_{j}\}$, satisfying the canonical
commutation relations\footnote{
For the $A$-type theory, it is customary to consider $A_{r-1}$ and to
embed all the roots in $\mathbb{R}^{r}$. This is accompanied by the
introduction of one more degree of freedom, $q_{r}$ and $p_{r}$.
The genuine $A_{r-1}$ theory corresponds to the relative coordinates and
their momenta, and the extra degree of freedom is the center of mass
coordinate and its momentum.}:
\begin{equation}
  [q_{j},p_{k}]=i\delta_{jk},\qquad [q_{j},q_{k}]=
  [p_{j},p_{k}]=0,\quad j,k=1,\ldots,r.
\end{equation}
These will be denoted by vectors in $\mathbb{R}^r$,
\begin{equation}
  q\eqdef (q_{1},\ldots,q_{r}),\quad
  p\eqdef (p_{1},\ldots,p_{r}),\quad 
  p\cdot q\eqdef \sum_{j=1}^r p_jq_j,\quad
  p^2\eqdef \sum_{j=1}^r p_j^2,\quad
  q^2\eqdef \sum_{j=1}^r q_j^2.
\end{equation}
The momentum operator $p_j$ acts as a differential operator
\[
  p_j=-i\frac{\partial}{\partial q_j}, \quad j=1,\ldots,r.
\]
The `factorised' Hamiltonian is
\begin{align}
  \mathcal{H}(p,q)&\eqdef\frac12\sum_{j=1}^r
  \Bigl(p_j-i\frac{\partial W}{\partial q_j}\Bigr)
  \Bigl(p_j+i\frac{\partial W}{\partial q_j}\Bigr),
  \label{facMHamiltonian}\\
  &=\frac12p^{2}+\frac{\omega^2}{2}q^2
  +\frac12\sum_{\rho\in\Delta_+}
  \frac{g_{|\rho|}(g_{|\rho|}-1)|\rho|^{2}}{(\rho\cdot q)^2}-{E}_0,
  \label{qCMHamiltonian}\\
  W(q)&\eqdef -\frac{\omega}{2}q^2
  +\sum_{\rho\in\Delta_+}g_{|\rho|}\log|\rho\cdot q|, \quad
  g_{|\rho|}>0,\quad \omega>0.
\end{align}
The summation is over the set of positive roots $\Delta_+$, with
$\Delta=\Delta_+\cup\,(-\Delta_+)$.
The real {\em positive\/} coupling constants \(g_{|\rho|}\) are defined
on orbits of the corresponding Coxeter group, {\it i.e.} they are
identical for roots of the same length.
The Hamiltonian is invariant under the finite reflection group (Coxeter
group, or Weyl group) generated by the set of roots $\Delta$:
\begin{equation}
  {\cal H}(s_{\alpha}(p),s_{\alpha}(q))={\cal H}(p,q), \quad
  \forall\alpha\in\Delta,
  \label{HamCoxinv}
\end{equation}
with the reflection \(s_{\alpha}\) defined by
\begin{equation}
  s_{\alpha}(x)\eqdef x-(\alpha^{\!\vee}\!\cdot x)\alpha, \quad
  \forall x\in\mathbb{R}^r,\quad
  \alpha^{\!\vee}\eqdef 2\alpha/|\alpha|^{2}.
  \label{Root_reflection}
\end{equation}

Obviously the Heisenberg equations of motion for the coordinates
$\{q_j\}$ are trivial
\begin{equation}
  i[\mathcal{H},q_j]=\frac{d}{dt}q_j=p_j,\quad j=1,\ldots,r,
\end{equation}
whereas those for the canonical momenta $\{p_j\}$ have the same form as
the Newton equations:
\begin{equation}
  \frac{d^2}{dt^2}q_j=\frac{d}{dt}p_j=i[\mathcal{H},p_j]=
  -\omega^2 q_j+\sum_{\rho\in\Delta_+}
  \frac{g_{|\rho|}(g_{|\rho|}-1) |\rho|^{2}\rho_j}{(\rho\cdot q)^3}, \quad
  j=1,\ldots,r.
\end{equation}
The hard repulsive potential $\sim{1/{(\rho\cdot q)^2}}$ near the
reflection hyperplane $H_{\rho}\eqdef\{q\in\mathbb{R}^r|\,\rho\cdot q=0\}$
is insurmountable at the quantum level as well as the classical.
Thus the motion is always confined within one Weyl chamber.
This feature allows us to constrain the configuration space to
the principal Weyl chamber ($\Pi$: set of simple roots)
\begin{equation}
  PW\eqdef\{q\in\mathbb{R}^r|\ \alpha\cdot q>0,\,\,\,\alpha\in\Pi\},
  \label{PW}
\end{equation}
without loss of generality.

The positive semi-definite form of the factorised Hamiltonian
\eqref{facMHamiltonian} simply allows the determination of the ground
state wavefunction:
\begin{equation}
  \mathcal{H}\phi_0(q)=0,\qquad
  \phi_0(q)\eqdef e^{W(q)}=\prod_{\rho\in\Delta_+}
  |\rho\cdot q|^{g_{|\rho|}}\cdot e^{-\frac{\omega}{2}q^2},
  \label{grsol}
\end{equation}
which is real and obviously square integrable
\begin{equation}
  \int_{PW}\phi_0^2(q)\,d^r\!q<\infty.
\end{equation}
The constant part ${E}_0$ of the Hamiltonian \eqref{facMHamiltonian}
is usually called the {\em ground state energy\/}
\begin{equation}
  {E}_0\eqdef\omega\Bigl(\frac{r}{2}+\sum_{\rho\in\Delta_+}g_{|\rho|}\Bigr).
\end{equation}

The excited energy spectrum is {\em integer spaced\/} and is
{\em independent of the coupling constants\/} $\{g_{|\rho|}\}$ \cite{kps}:
\begin{alignat}{2}
  \mathcal{H}\phi_{\bf n}(q)&=\mathcal{E}_{\bf n}\phi_{\bf n}(q),&\quad
  {\bf n}&\eqdef (n_1,\ldots,n_r),\quad
  n_j\in\mathbb{N}\eqdef\mathbb{Z}_{\geq 0},\\
  \mathcal{E}_{\bf n}&\eqdef\omega N_{\bf n},&
  N_{\bf n}&\eqdef\sum_{j=1}^r n_j\,f_j.
\end{alignat}
Here $\{f_j\}$ are the integers related to the {\em exponents\/}
$\{e_j\}$ of $\Delta$:
\begin{equation}
  f_j\eqdef 1+e_j.
\end{equation}
They indicate the degrees where independent Coxeter invariant
polynomials exist. The set of integers
\begin{equation}
  F_\Delta\eqdef \{f_1,f_2,\ldots,f_r\}
\end{equation}
is shown in Table I for each root system $\Delta$.
\begin{center}
\begin{tabular}{||c|c||c|c||}
  \hline
  \(\Delta\)& \(F_\Delta\) &\(\Delta\)& \(F_\Delta\)\\
  \hline
  \(A_{r-1}\) & \(\{2,3,\ldots,r,1\}\) & \(E_8\) &
  \(\{2,8,12,14,18,20,24,30\}\)\\
  \hline
  \(B_r\) & \(\{2,4,6,\ldots,2r\}\) & \(F_4\) & \(\{2,6,8,12\}\) \\
  \hline
  \(C_r\) & \(\{2,4,6,\ldots,2r\}\) & \(G_2\) & \(\{2,6\}\) \\
  \hline
  \(D_r\) & \(\{2,4,\ldots,2r-2,r\}\) & \(I_2(m)\) & \(\{2,m\}\) \\
  \hline
  \(E_6\) & \(\{2,5,6,8,9,12\}\) & \(H_3\) & \(\{2,6,10\}\) \\
  \hline
  \(E_7\) & \(\{2,6,8,10,12,14,18\}\) & \(H_4\) & \(\{2,12,20,30\}\) \\
  \hline
\end{tabular}\\
\bigskip
Table I: The set of integers \(F_\Delta=\{f_1,f_2,\ldots,f_r\}\) for
which independent\\
Coxeter invariant polynomials exist.\footnote{
For $A_{r-1}$ root system, $f_r=1$ corresponds to the degree of freedom
for the center of mass coordinate. The $B_r$ and $C_r$ Calogero 
systems are equivalent.
}
\end{center}
Excited states eigenfunctions have the following general structure:
\begin{equation}
  \phi_{\bf n}(q)=\phi_0(q)P_{\bf n}(q),
  \label{polyPdef}
\end{equation}
in which $P_{\bf n}(q)$ is a Coxeter invariant polynomial in
$\{q_j\}$ of degree $N_{\bf n}$.

\subsection{Quantum Lax Pair}
\label{lax}

The derivation of the Heisenberg operator solutions depends heavily on
the universal Lax pair which applies to any root system. The universal
Lax pair operators \cite{bms,kps} are
\begin{align}
  L(p,q)&\eqdef p\cdot\hat{H}+X(q),\qquad 
  X(q)\eqdef i\sum_{\rho\in\Delta_{+}}g_{|\rho|}\,
  \frac{(\rho\cdot\hat{H})}{\rho\cdot q}\hat{s}_{\rho},
  \label{LaxOpDef}\\
  M(q)&\eqdef -\frac{i}{2}\sum_{\rho\in\Delta_{+}}g_{|\rho|}
  \frac{|\rho|^2}{(\rho\cdot q)^2}\,\hat{s}_{\rho}
  +\frac{i}{2}\sum_{\rho\in\Delta_{+}}g_{|\rho|}
  \frac{|\rho|^2}{(\rho\cdot q)^2}\,\times I,
  \label{Mtildef}
\end{align}
in which $I$ is the identity operator and
$\{\hat{s}_{\alpha}:\,\alpha\in\Delta\}$ are the reflection operators
of the root system. They act on a set of $\mathbb{R}^{r}$ vectors
$\mathcal{R}\eqdef\{\mu^{(k)}\!\in\mathbb{R}^r|\,k=1,\ldots,d\,\}$,
permuting them under the action of the reflection group.
The vectors in $\mathcal{R}$ form a basis for the representation space 
${\bf V}$ of dimension $d$.
The operator $M$ satisfies the relation \cite{bms,kps}
\begin{equation}
  \sum_{\mu\in\mathcal{R}}M_{\mu\nu}=\sum_{\nu\in\mathcal{R}}M_{\mu\nu}=0,
  \label{sumMzero}
\end{equation}
which is essential for deriving quantum conserved quantities and
annihilation-creation operators. The matrix elements of the operators
$\{\hat{s}_{\alpha}:\,\alpha\in\Delta\}$ and 
$\{\hat{H}_{j}:\,j=1,\ldots,r\}$ are defined as follows:
\begin{equation}
  (\hat{s}_{\rho})_{\mu\nu}\eqdef\delta_{\mu,s_\rho(\nu)}=
  \delta_{\nu,s_\rho(\mu)}, \quad
  (\hat{H}_{j})_{\mu\nu}\eqdef\mu_j\delta_{\mu\nu},\quad 
  \rho\in\Delta, \quad  \mu, \nu\in\mathcal{R}.
  \label{sHdef}
\end{equation}

The Lax equation
\begin{equation}
  i[\mathcal{H},L]=\frac{d}{dt}L=[L,M]
  \label{laxeq0}
\end{equation}
is equivalent to the Heisenberg equation of motion for $\{q_j\}$ and
$\{p_j\}$ for the Hamiltonian \eqref{qCMHamiltonian} without the
harmonic confining potential, {\em i.e,\/} $\omega=0$.
It should be emphasised that the l.h.s. of \eqref{laxeq0} is a quantum 
commutator, whereas the r.h.s. is a matrix commutator as well as quantum.
The full Heisenberg equations of motion with the harmonic confining
potential read
\begin{equation}
  i[{\cal H},L^{\pm}]=\frac{d}{dt}{L^{\pm}}
  =[L^{\pm},{M}]\pm i\omega L^{\pm},
  \label{omegaLM}
\end{equation}
in which $M$ is the same as before \eqref{Mtildef}, and $L^{\pm}$ and
$Q$ are defined by
\begin{equation}
  L^{\pm}\eqdef L\pm i\omega Q, \quad Q\eqdef q\cdot\hat{H},
\quad (L^+)^\dagger=L^-,
  \label{Lpmdef}
\end{equation}
with $L$, $\hat{H}$ as earlier \eqref{LaxOpDef}, \eqref{sHdef}.
One direct consequence of the Lax pair equation \eqref{omegaLM} is the
existence of a wide variety of {\em quantum conserved quantities\/}:
\begin{equation}
  \frac{d}{dt}\text{Ts}(L^{\epsilon_1}L^{\epsilon_2}\cdots L^{\epsilon_k})=0,
  \quad
  \epsilon_j\in\{+,-\},\quad \forall k\in 2\mathbb{N},\quad
  \sum_{j=1}^k\epsilon_j=0,
\end{equation}
in which $\text{Ts}(A)$ denotes the {\em total sum\/} of a matrix $A$
\cite{bms,kps} with suffices $\mu,\nu\in\mathcal{R}$:
\begin{equation}
  \text{Ts}(A)\eqdef \sum_{\mu,\nu\in\mathcal{R}}A_{\mu\,\nu}.
\end{equation}
This is a simple outcome of the property \eqref{sumMzero} of the matrix $M$
\cite{bms,kps,shas,ujiwa}.
The quantum conserved quantities are the simplest example of more general
results:
\begin{alignat}{2}
  i[{\cal H},\text{Ts}(L^{\epsilon_1}L^{\epsilon_2}\cdots L^{\epsilon_k})]
  &=\frac{d}{dt}\text{Ts}
  (L^{\epsilon_1}L^{\epsilon_2}\cdots L^{\epsilon_k}),
  &\qquad& \forall k\in\mathbb{N},\n
  &=i\omega m\, \text{Ts}(L^{\epsilon_1}L^{\epsilon_2}\cdots L^{\epsilon_k}),
  && \sum_{j=1}^k\epsilon_j=m.
\end{alignat}
In other words, 
$\text{Ts}(L^{\epsilon_1}L^{\epsilon_2}\cdots L^{\epsilon_k})$ shifts
the eigenvalue of $\mathcal{H}$ by $m$ unit (or, $\omega m$) when it
acts on any eigenstate of $\mathcal{H}$. Namely such operators are all
candidates of {\em annihilation-creation operators} for $\mathcal{H}$:
\begin{equation}
  e^{i\mathcal{H}t}\,\text{Ts}
  (L^{\epsilon_1}L^{\epsilon_2}\cdots L^{\epsilon_k})e^{-i\mathcal{H}t}
  =\text{Ts}(L^{\epsilon_1}L^{\epsilon_2}\cdots L^{\epsilon_k})\,
  e^{i\omega mt},\qquad
  \sum_{j=1}^k\epsilon_j=m.
  \label{prodform}
\end{equation}
Note that the order of $L^+$ and $L^-$ is immaterial.

The results and statements in this subsection are {\em universal\/},
{\em i.e.\/} they are valid for any root system $\Delta$ and for any
choice of the Lax pair matrices, {\em i.e.\/} the choice of $\mathcal{R}$.
For deriving the explicit forms of the Heisenberg operator solutions,
however, it is convenient to choose as $\mathcal{R}$ the set of the least
dimensions for each $\Delta$.
For the classical root systems, $A$, $B$ 
and $D$, they are the set of vector weights\footnote{
To be more precise, for $B_r$ they should be called the set of short
roots.}:
\begin{alignat}{3}
  A_{r-1}\ \ \,&:&\quad \mathcal{R}&\eqdef 
  \{{\bf e}_1,{\bf e}_2,\ldots,{\bf e}_{r}\},&\quad
  d&\eqdef \#\mathcal{R}=r,\\
  B_r,\,D_r&:& \mathcal{R}&\eqdef
  \{\pm{\bf e}_1,\pm{\bf e}_2,\ldots,\pm{\bf e}_{r}\},&
  d&\eqdef\#\mathcal{R}=2r,
  \label{bcdvec}
\end{alignat}
in which $\{{\bf e}_j\}$ are the orthonormal basis of $\mathbb{R}^r$, 
${\bf e}_j\cdot {\bf e}_k=\delta_{j\,k}$.
We list in the Appendix the preferred choice of $\mathcal{R}$ for the
exceptional and the non-crystallographic root systems.

\section{Exact Heisenberg Solutions}
\setcounter{equation}{0}

Now let us proceed to derive the explicit Heisenberg operator solutions
for various `sinusoidal coordinates' \cite{os7,nieto} as many as the
degree of freedom. For simplicity of presentation, let us fix the
angular frequency of the harmonic confining potential as unity, 
$\omega=1$ hereafter. We have already known one exact Heisenberg
operator solution. In \cite{os7} it was shown that the quadratic invariant
\begin{equation}
  \eta^{(1)}\propto q^2=\sum_{j=1}^r q_j^2
\end{equation}
is the simplest sinusoidal coordinate:
\begin{equation}
  [\mathcal{H},[\mathcal{H},\eta^{(1)}]]
  =4(\eta^{(1)}-c_1\mathcal{H}-c_2), \quad 
  c_1,c_2:\, \text{const},
\end{equation}
where $c_1$ is given by $\eta^{(1)}=c_1q^2$ and $c_2=c_1E_0$.
As shown in \cite{os7} the exact Heisenberg solution
\begin{equation}
  e^{i\mathcal{H}t}\,\eta^{(1)}\,e^{-i\mathcal{H}t}
\end{equation}
is easily evaluated. It should be noted that any root system has a
degree 2 Coxeter invariant $2\in F_\Delta$ (Table I), which is
proportional to $\eta^{(1)}$. In fact, $\eta^{(1)}$ is a sinusoidal
coordinate \cite{os7} for a Hamiltonian $\mathcal{H}_{\text{ge}}$
more general than the Calogero system;
a general homogeneous potential of degree $-2$ within a confining harmonic
potential \cite{gamba}:
\begin{equation}
  \mathcal{H}_{\text{ge}}=\frac{1}{2}\sum_{j=1}^r(p_j^2+q_j^2)+V(q),\quad
  \sum_{j=1}^r q_j\frac{\partial}{\partial q_j}\,V(q)=-2\,V(q),
\end{equation}
which satisfies $[\mathcal{H}_{\text{ge}},[\mathcal{H}_{\text{ge}},q^2]]
=4(q^2-\mathcal{H}_{\text{ge}})$.

For the Calogero system based on any root system $\Delta$,
we will derive the explicit forms of the Heisenberg operator solutions:
\begin{equation}
  e^{i\mathcal{H}t}\,\eta^{(j)}\,e^{-i\mathcal{H}t},\quad j=1,\ldots, r,
\end{equation}
for a complete set of sinusoidal coordinates defined by
\begin{eqnarray}
  \{\eta^{(1)},\eta^{(2)},\ldots,\eta^{(r)}\},\quad
  \eta^{(j)}\eqdef \mbox{Ts}(Q^{f_j})=\mbox{Tr}(Q^{f_j}),\quad
  f_j\in F_\Delta.
  \label{etadefs}
\end{eqnarray}
In \eqref{etadefs}, the matrix $Q$ \eqref{Lpmdef} is diagonal therefore its
total sum (Ts) is the same as the trace (Tr).
Let us note that any Coxeter invariant polynomial in $\{q_j\}$ can be
expressed as a polynomial in $\{\eta^{(1)},\eta^{(2)},\ldots,\eta^{(r)}\}$.
It is easy to verify $\eta^{(1)}\propto q^2$ for an root system.
The higher sinusoidal coordinates for the classical root systems are:
\begin{alignat}{3}
  A_{r-1}\ \ \,&:&\quad \eta^{(k)}&\propto \sum_{j=1}^{r}q_j^{k+1},&\quad 
  k&=1,\ldots,r-1;\quad \eta^{(r)}\propto \sum_{j=1}^{r}q_j, \\
  B_r,\,D_r&:& \eta^{(k)}&\propto \sum_{j=1}^{r}q_j^{2k},&
  k&=1,\ldots,r,
\end{alignat}
except for $\eta^{(r)}$ for $D_r$ which reads
\begin{equation}
  \eta^{(r)}\propto \prod_{j=1}^{r}q_j.
\end{equation}
See the remark at the end of the section.
As shown in Table I, all the integers $\{f_j\}$ are {\em even\/} for
$B_r$ and $D_r$, except for $f_r=r$ for {\em odd\/} $r$ of
$D_r$. This is also related to the fact that
\[
  \text{Tr}(Q^{2l+1})=\sum_{j=1}^rq_j^{2l+1}+\sum_{j=1}^r(-q_j)^{2l+1}=0,
\]
for the Lax operators based on the vector weights \eqref{bcdvec}.

The sinusoidal coordinates take various different forms for the exceptional
and non-crystallographic root systems.
Note that the overall normalisation of $\{\eta^{(k)}\}$ is immaterial.

\bigskip
The derivation of exact Heisenberg operator solutions is quite elementary.
Let us introduce a {\em generating function\/} of the total sum of the 
homogeneous polynomials in $L^+$ and $L^-$:
\begin{equation}
  G^{(j)}(s)\eqdef \text{Ts}\bigl((L^++s\,L^-)^{f_j}\bigr),\quad 
  s\in\mathbb{C},\quad f_j\in F_\Delta,
\end{equation}
which is a polynomial in $s$ of degree $f_j$
\begin{equation}
  G^{(j)}(s)=\sum_{l=0}^{f_j}b_{f_j;f_j-2l}\,s^{l},
\qquad b_{f_j;l}^\dagger=b_{f_j;-l}.
\end{equation}
The coefficient $b_{f_j;f_j-2l}$ is the total sum of a
{\em completely symmetric product} consisting of $f_j-l$ times $L^+$ and
$l$ times $L^-$ and it can be explicitly evaluated.
As shown in \eqref{prodform}, we obtain
\begin{equation}
  e^{i\mathcal{H}t}\,b_{f_j;f_j-2l}\,e^{-i\mathcal{H}t}=
  b_{f_j;f_j-2l}\,e^{i(f_j-2l)t}.
\end{equation}
Namely $b_{f_j;f_j-2l}$ is either an {\em annihilation operator\/}
($l>f_j/2$) being the {\em positive energy part\/} or a
{\em creation operator\/} ($l<f_j/2$) being the
{\em negative energy part\/} or a conserved quantity ($l=f_j/2$),
the constant part. 
The conserved quantities are in general not in involution
\begin{equation}
  [b_{f_j;\,0},b_{f_k;\,0}]\neq0,\quad j\neq k.
\end{equation}

On the other hand, for $s=-1$ we obtain
\[
  L^+-L^-=2iQ
\]
and
\begin{align}
  G^{(j)}(-1)&=(2i)^{f_j}\text{Ts}(Q^{f_j})=(2i)^{f_j}\eta^{(j)}
  =\sum_{l=0}^{f_j}(-1)^l\,b_{f_j;f_j-2l},\n
  \eta^{(j)}&=(2i)^{-f_j}\sum_{l=0}^{f_j}(-1)^l\,b_{f_j;f_j-2l}.
\end{align}
Thus we arrive at the main result of the paper;
{\em the complete set of exact Heisenberg operator solutions\/} for
the `sinusoidal coordinates' $\{\eta^{(j)}\}$:
\begin{equation}
  \eta^{(j)}(t)\eqdef e^{i\mathcal{H}t}\,\eta^{(j)}\,e^{-i\mathcal{H}t}
  =(2i)^{-f_j}\sum_{l=0}^{f_j}(-1)^l\,b_{f_j;f_j-2l}\,e^{i(f_j-2l)t},\quad
  j=1,2,\ldots,r.
  \label{Heisensol}
\end{equation}
This clearly shows that $\eta^{(j)}(t)$ is a superposition of various
sinusoidal motions. The trivial fact that these `sinusoidal coordinates'
commute among themselves
\begin{equation}
  \eta^{(j)}\eta^{(k)}=\eta^{(k)}\eta^{(j)},\quad j\neq k=1,\ldots, r
\end{equation}
is translated into the commutation relations among the 
annihilation-creation operators
\begin{equation}
  \sum_{\genfrac{}{}{0pt}{}{l,m}{l+m:\text{fixed}}}
  [b_{f_j;f_j-2l},b_{f_k;f_k-2m}]=0.
  \label{an-crcomm}
\end{equation}
Among the annihilation-creation operators belonging to $\eta^{(j)}$,
the two extreme ones corresponding to $l=0$ and $l=f_j$ have a special
meaning. They consist of $L^+$ ($L^-$) only
\begin{equation}
  b_{f_j;f_j}=\text{Ts}\bigl((L^+)^{f_j}\bigr),\quad
  b_{f_j;-f_j}=\text{Ts}\bigl((L^-)^{f_j}\bigr),\quad
  b_{f_j;f_j}^\dagger=b_{f_j;-f_j}
\end{equation}
and commute among themselves
\begin{equation}
  [b_{f_j;f_j},b_{f_k;f_k}]=0,\quad
  [b_{f_j;-f_j},b_{f_k;-f_k}]=0,\quad j,k=1,\ldots,r,
\end{equation}
as is clear from \eqref{an-crcomm}.
These special annihilation-creation operators have been known for
some time \cite{kps,OP1,ujiwa,Br}.

Since the $l$-th term in \eqref{Heisensol} is annihilated by
$d/dt-i(f_j-2l)$, we obtain a linear differential equation with
constant coefficients satisfied by $\eta^{(j)}(t)$:
\begin{equation}
  \prod_{\genfrac{}{}{0pt}{}{l=-f_j}{\text{step 2}}}^{f_j}
  \Bigl(\frac{d}{dt}-il\Bigr)\cdot\eta^{(j)}(t)=0.
\end{equation}
Equivalently this can be rewritten as
\begin{equation}
  \prod_{\genfrac{}{}{0pt}{}{l=-f_j}{\text{step 2}}}^{f_j}
  \bigl(\text{ad}(\mathcal{H})-l\bigr)\cdot\eta^{(j)}=0.
  \label{closure}
\end{equation}
Here $\mbox{ad}(\mathcal{H})$ denotes a commutator
$\mbox{ad}(\mathcal{H})X\eqdef [\mathcal{H},X]$ for any operator $X$.
For even $f_j$ case, the factor $d/dt-i(f_j-2l)$ for $l=f_j/2$ could be
omitted as it annihilates the conserved quantity. Then we obtain a
differential equation of order $f_j$
\begin{align}
  \prod_{\genfrac{}{}{0pt}{}{l=-f_j,\neq 0}{\text{step 2}}}^{f_j}
  \Bigl(\frac{d}{dt}-il\Bigr)\cdot
  \bigl(\eta^{(j)}(t)-2^{-f_j}b_{f_j;\,0}\bigr)&=0,\\
  \prod_{\genfrac{}{}{0pt}{}{l=-f_j,\neq 0}{\text{step 2}}}^{f_j}
  \bigl(\text{ad}(\mathcal{H})-l\bigr)\cdot
  \bigl(\eta^{(j)}-2^{-f_j}b_{f_j;\,0}\bigr)&=0,
  \label{closure2}
\end{align}
instead of $f_j+1$ for odd $f_j$ case.
In \cite{os7}, for a wide class of solvable quantum systems with one
degree of freedom, we discussed the `closure relation'
\begin{equation}
  [\mathcal{H},[\mathcal{H},\eta]]=\eta R_0(\mathcal{H})
  +[\mathcal{H},\eta]R_1(\mathcal{H})+R_{-1}(\mathcal{H})
\end{equation}
with $R_i(\mathcal{H})^{\dagger}=R_i(\mathcal{H})$.
By introducing $\alpha_{\pm}(\mathcal{H})$ as
$R_1(\mathcal{H})=\alpha_+(\mathcal{H})+\alpha_-(\mathcal{H})$ and
$R_0(\mathcal{H})=-\alpha_+(\mathcal{H})\alpha_-(\mathcal{H})$,
this closure relation is rewritten as
\begin{equation}
  \bigl(\text{ad}(\mathcal{H})+\alpha_+(\mathcal{H})\bigr)
  \bigl(\text{ad}(\mathcal{H})+\alpha_-(\mathcal{H})\bigr)
  \bigl(\eta+R_0(\mathcal{H})^{-1}R_{-1}(\mathcal{H})\bigr)=0.
\end{equation}
Eqs. \eqref{closure} and \eqref{closure2} are multi-particle
generalisation of this relation.

When a sinusoidal coordinate $\eta^{(j)}$ is multiplied to an
eigenvector $\phi_{\bf n}$ of $\mathcal{H}$,
\begin{equation}
  (2i)^{f_j}\eta^{(j)}\phi_{\bf n}
  =\sum_{l=0}^{f_j}(-1)^l b_{f_j;f_j-2l}\,\phi_{\bf n},
\end{equation}
the $l$-th term belongs to the eigenspace of $\mathcal{H}$ with the
eigenvalue $\mathcal{E}_{\bf n}+f_j-2l$. Thus these $f_j+1$ terms are all
orthogonal to each other.
This is a {\em multi-variable generalisation\/} of the
{\em three-term recursion relation\/} of the orthogonal polynomials of
one variable \cite{os7}.
Through a similarity transformation in terms of the ground state
wavefunction $\phi_0(q)$ \eqref{grsol}, let us define
\begin{align}
  \tilde{L}&\eqdef\phi_0^{-1}\circ L \circ \phi_0,\qquad
  \tilde{L}^\pm\eqdef\phi_0^{-1}\circ\tilde{L}^{\pm}\circ\phi_0
  =\tilde{L}\pm i Q,\\
  \tilde{b}_{f_j;f_j-2l}&\eqdef
  \phi_0^{-1}\circ {b}_{f_j;f_j-2l}\circ \phi_0, \quad
  j=1,\ldots,r,\quad l=0,1,\ldots,f_j.
\end{align}
Needless to say the identity
\[
  \phi_0^{-1}\circ\text{Ts}(A^n)\circ\phi_0
  =\text{Ts}\bigl((\phi_0^{-1}\circ A\circ\phi_0)^n\bigr)
\]
holds for any matrix $A$ consisting of operators.
Then one can present the corresponding results for the multi-variable
orthogonal polynomials $\{P_{\bf n}(q)\}$ \eqref{polyPdef} constituting
the eigenvectors:
\begin{eqnarray}
  (2i)^{f_j}\eta^{(j)}P_{\bf n}(q)=\sum_{l=0}^{f_j}(-1)^l 
  \tilde{b}_{f_j;f_j-2l}\,P_{\bf n}(q).
\end{eqnarray}
The operators $\tilde{L}$, $\tilde{L}^\pm$ and $\tilde{b}_{f_j;f_j-2l}$
are closely related to the Dunkl operators \cite{kps,Dunk}.

\bigskip
The annihilation-creation operators provide an algebraic solution method
of the Calogero systems. The entire Hilbert space is generated by the
multiple application of creation operators on the ground state
wavefunction  $\phi_0(q)=e^{W(q)}$:
\begin{equation}
  \prod_{j=1}^r
  \prod_{\genfrac{}{}{0pt}{}{0<l\leq f_j}{l\equiv f_j(\text{mod}\,2)}}
  b_{f_j;l}^{n_{(j,l)}}\cdot\phi_0,\quad
  \forall n_{(j,l)} \in\mathbb{N},
  \label{generalstates}
\end{equation}
whereas $\phi_0$ is destroyed by all the annihilation operators
\begin{equation}
  b_{f_j;l}\phi_0=0,\quad -f_j\leq l<0,\ \ l\equiv f_j (\text{mod}\,2).
\end{equation}
Obviously the above states \eqref{generalstates} are over-complete and
the orthogonality of various eigenvectors belonging to the same
degenerate eigenspace is not guaranteed.
For example, a complete basis of the Hilbert space is given by
using $b_{f_j;f_j}$ only \cite{kps}
\begin{equation}
  \prod_{j=1}^r b_{f_j;f_j}^{n_j}\cdot\phi_0,\quad
  \forall n_j\in\mathbb{N}.
\end{equation}

\bigskip
\paragraph{Remark on $D$-type Theory}

As noticed above, the $D$-type theory requires separate treatment due
to the special $f_r=r$. We need Lax pairs based on the {\em spinor\/}
and {\em anti-spinor\/} weights:
\begin{eqnarray}
  \mathcal{R}\eqdef \Bigl\{\pm\frac{1}{2}{\bf e}_1\pm\frac{1}{2}{\bf e}_2
  \pm\cdots\pm\frac{1}{2}{\bf e}_{r}\Bigr\},\quad
  d\eqdef\#\mathcal{R}=2^{r-1}.
\end{eqnarray}
Here the the number of $-$ signs is {\em even\/} ({\em odd\/}) for
the spinor (anti-spinor) weights.
They both form $2^{r-1}$ dimensional representations of the Lie-algebra
$D_r$, and these representations are called {\em minimal\/}.
The minimal weights for the $A$, $D$ and $E$ root systems have played an
important role in constructing simple Lax pair representations \cite{bcs1}.

For $D_{\text{odd}}$ we simply use the Lax matrix  $L^{\pm}_{(\text{sp})}$
($L^{\pm}_{(\text{as})}$) based on either the spinor (sp) or the
anti-spinor (as) weights and proceed in the same way as above:
\begin{align}
  G^{(r)}_{(\text{sp})}(s)&\eqdef
  \text{Ts}\bigl((L^+_{(\text{sp})}+s\,L^-_{(\text{sp})})^r\bigr),\quad
  s\in\mathbb{C},\quad f_r=r,\n
  &=\sum_{l=0}^r b_{(\text{sp})\,r;r-2l}\,s^{l}.
\end{align}
Then we obtain the expansion of $\eta^{(r)}$ in terms of the
annihilation-creation operators and the corresponding exact Heisenberg
operator solution
\begin{align}
  G^{(r)}_{(\text{sp})}(-1)&=
  (2i)^{r}\text{Ts}(Q_{(\text{sp})}^{r})=(2i)^{r}\eta^{(r)}
  =\sum_{l=0}^{r}(-1)^l\,b_{(\text{sp})\,r;r-2l},\\
  \eta^{(r)}&=(2i)^{-r}\sum_{l=0}^{r}(-1)^l\,b_{(\text{sp})\,r;r-2l}
  \propto q_1q_2\cdots q_r,
  \label{etarsp}\\
  \eta^{(r)}(t)&\eqdef e^{i\mathcal{H}t}\,\eta^{(r)}\,e^{-i\mathcal{H}t}
  =(2i)^{-r}\sum_{l=0}^{r}(-1)^l\,b_{(\text{sp})\,r;r-2l}\,e^{i(r-2l)t}.
\end{align}

For $D_{\text{even}}$, the situation is slightly more complicated,
partly because of the existence of another sinusoidal coordinate
$\eta^{(r/2)}$, which is a Coxeter invariant polynomial of $\{q_j\}$ of
degree $r$, too.
Here we prepare the Lax matrices based on both the spinor
$L^{\pm}_{(\text{sp})}$ and the anti-spinor weights
$L^{\pm}_{(\text{as})}$, since
\begin{equation}
  \eta^{(r)}\propto
  \text{Ts}(Q_{(\text{sp})}^{r})-\text{Ts}(Q_{(\text{as})}^{r})
  \propto q_1q_2\cdots q_r.
  \label{etarsp-as}
\end{equation}
It is quite elementary to verify \eqref{etarsp} and \eqref{etarsp-as}.
Let us introduce two `generating functions'
\begin{align}
  G^{(r)}_{(\text{sp})}(s)&\eqdef
  \text{Ts}\bigl((L^+_{(\text{sp})}+s\,L^-_{(\text{sp})})^{r}\bigr)
  =\sum_{l=0}^{r}b_{(\text{sp})\,r;r-2l}^{}\,s^{l},\\
  G^{(r)}_{(\text{as})}(s)&\eqdef
  \text{Ts}\bigl((L^+_{(\text{as})}+s\,L^-_{(\text{as})})^{r}\bigr)
  =\sum_{l=0}^{r}b_{(\text{as})\,r;r-2l}^{}\,s^{l}.
\end{align}
Next we introduce the sinusoidal coordinate $\eta^{(r)}$ as their
difference at $s=-1$,
\begin{align}
  \eta^{(r)}&\eqdef
  (2i)^{-r}\bigl(G^{(r)}_{(\text{sp})}(-1)-G^{(r)}_{(\text{as})}(-1)\bigr)
  =\text{Ts}(Q_{(\text{sp})}^{r})-\text{Ts}(Q_{(\text{as})}^{r})\n
  &=(2i)^{-r}\sum_{l=0}^{r}(-1)^l\,
  \bigl(b_{(\text{sp})\,r;r-2l}-b_{(\text{as})\,r;r-2l}\bigr).
\end{align}
We obtain the corresponding exact Heisenberg operator solution
\begin{equation}
  \eta^{(r)}(t)\eqdef e^{i\mathcal{H}t}\,\eta^{(r)}\,e^{-i\mathcal{H}t}
  =(2i)^{-r}\sum_{l=0}^{r}(-1)^l
  \bigl(b_{(\text{sp})\,r;r-2l}-b_{(\text{as})\,r;r-2l}\bigr)e^{i(r-2l)t}.
\end{equation}
This completes the derivation of the exact Heisenberg operator solutions
for the $D$-type Calogero systems.

\section{Summary and Comments}
\label{comments}
\setcounter{equation}{0}

A complete set of exact Heisenberg operator solutions, as many as the
degree of freedom, is constructed for the Calogero systems based on any
root system, including the exceptional and non-crystallographic ones.
Based on the complete set, one can write down the Heisenberg operator
solution $e^{i\mathcal{H}t}\,A\,e^{-i\mathcal{H}t}$ for any operator
$A$ expressible as a polynomial in the sinusoidal coordinates
$\{\eta^{(j)}\}$. This is the first demonstration of the {\em exact
solvability of multi-particle quantum mechanics\/} in the
{\em Heisenberg picture\/}.
At the same time, these Heisenberg operator solutions provide the
explicit forms of various {\em annihilation-creation operators\/},
as the {\em positive\/} and {\em negative energy parts\/}.
Their commutation relations are, in general, quite involved.
As in the simplest case of degree one quantum mechanics \cite{os7},
these sinusoidal coordinates and their expansion into the
annihilation-creation operators provide the explicit forms of the
{\em multi-variable generalisation of the three term recursion
relations\/} for the orthogonal polynomials constituting the
multi-variable eigenfunctions.
The derivation of the Heisenberg operator solutions is a simple
consequence of the {\em universal\/} Lax pair, which manifests the
quantum integrability of Calogero systems based on {\em any root
system\/}.

Let us conclude this paper with a few comments on possible future
directions of the present research.
For better understanding of multi-particle quantum mechanics in general,
it is desirable to enlarge the list of exact Heisenberg operator
solutions. The obvious candidates are:
the Sutherland systems \cite{sut} having trigonometric potentials,
various super-symmetric generalisations of the Calogero-Sutherland
systems \cite{bms,shas}, the Ruijsenaars-Schneider-van-Diejen systems
\cite{RSvD} which are `discrete' counterparts of the Calogero-Sutherland
systems, and the (affine) Toda molecules.

The newly found annihilation-creation operators suggest an interesting
possibility of introducing {\em multi-particle coherent states\/} as
common eigenstates of certain annihilation operators.
It is a good challenge to construct explicit examples of such
multi-particle coherent states having mathematically elegant structure
and/or practical use.

A completely integrable system, including the Calogero-Sutherland-Moser
systems, has the so-called {\em hierarchy structure\/}.
It is characterised by the existence of mutually involutive conserved
quantities
\begin{equation}
  \mathcal{H}_1,\mathcal{H}_2,\ldots,\mathcal{H}_r,\quad
  [\mathcal{H}_j,\mathcal{H}_k]=0,\quad
  j,k=1,\ldots,r,
\end{equation}
which could be adopted as independent Hamiltonians generating different
but compatible time-flows; $t_1$, $t_2$,\ldots, $t_r$, as many as the
degree of freedom.
It is a good challenge to construct common Heisenberg operator solutions
to all the flows of the hierarchy
\begin{equation}
  e^{i\sum_{j=1}^r\mathcal{H}_j t_j}\,\tilde{\eta}^{(k)}
  e^{-i\sum_{j=1}^r\mathcal{H}_jt_j},\quad k=1,\ldots,r.
\end{equation}

\section*{Acknowledgements}

This work is supported in part by Grants-in-Aid for Scientific
Research from the Ministry of Education, Culture, Sports, Science and
Technology, No.18340061 and No.19540179.

\section*{Appendix: The Preferred Choice of $\mathcal{R}$}
\setcounter{equation}{0}
\label{append}
\renewcommand{\theequation}{A.\arabic{equation}}

Here we list, for the exceptional and the non-crystallographic root
systems, the set $\mathcal{R}$ to be used for the explicit evaluation of
the Heisenberg operator solutions. They are of the lowest dimensionality.
\begin{enumerate}
\item
\(E_6\):
The weights of {\bf 27} (or \({\bf \overline{27}}\)) dimensional
representation of the Lie algebra.
They are minimal representations.
\item
\(E_7\):
The weights of {\bf 56} dimensional representation of the Lie algebra.
This is a minimal representation.
\item
\(E_8\):
The set consisting of all 240 roots.
\item
\(F_4\):
Either of the set consisting of all 24 long roots or 24 short roots.
\item
\(G_2\):
Either of the set consisting of all 6 long roots or 6 short roots.
\item
\(I_2(m)\):
The set consisting of the vertices \(R_m\) of the regular $m$-gon
\begin{equation}
  R_m=\bigl\{v_j=\bigl(\cos(2k\pi/m+t_0),\sin(2k\pi/m+t_0)\bigr)
  \in\mathbb{R}^2\bigm|k=1,\ldots,m\bigr\},
  \label{vmpara}
\end{equation}
in which $t_0=0\ (\pi/2m)$ for $m$ even (odd).

\item
\(H_3\):
The set consisting of all 30 roots.
\item
\(H_4\):
The set consisting of all 120 roots.
\end{enumerate}


\end{document}